# On the Development of Testing Tool for a Satellite Gyro Sensor

Harry Septanto
Satellite Technology Center
National Institute of Aeronautics and Space (LAPAN)
Bogor, Indonesia

Desti Ika Suryanti
Satellite Technology Center
National Institute of Aeronautics and Space (LAPAN)
Bogor, Indonesia

***Abstract*— Attitude determination and control system (ADCS) in a satellite takes an important role to make sure that the satellite mission will be achieved. In the development phase, the ADCS is required to pass whole testing levels, including component level verification. As an important component in the ADCS, the gyro sensors must meet in that requirement. However, the testing tool for the component level test often rely on another satellite component. Since the testing line is not directly connected to the gyro, some failures between those components will be hard to be identified. This paper proposes a testing tool that operates without involves any other satellite components. The proposed testing tool consists of software and hardware part. The software part is a desktop application that generates trigger signal for requesting gyro measurement data and the hardware part is USB to RS-422 converter and USB cable for gyro power supply. Implementation and experimental results show that the proposed testing tool is promising to be a standard tool for component level verification of the satellite gyro sensor.**

***Keywords-gyro sensor; testing; software application; serial communication; satellite;***

## I. Introduction

In recent years, development of low mass and size satellites that bring wide variety mission are growing rapidly. In [1], Damé et al presents scientific mission of microsatellite that concerns to observe space environment and the onset of interplanetary coronal mass ejection. Therefore, risk mitigation effort due to extreme space weather impact on ground-based and Space-based infrastructure—e.g. telecommunication, banking, navigation—can be built. It was designed to use MYRIADE micro-satellite platform [2]. Yamagiwa et al [3] reports development of microsatellites for demonstration of space elevator technology demonstration. One of the microsatellites called STARS-C was launched in late 2016 via Kibo module on International Space Station. Its mission is for the tether deployment technology experiment. The other microsatellite called STARS-E is still developing that will bring mission to verify the climber operation in Space. Lucente et al [4] reports research result with the purpose of implement a low-cost W-band experiment on-board as a nanosatellite mission. The mission objectives including to obtain W-band troposphere attenuation measurement and to validate the W-band hardware in space. All of these satellites requires an attitude determination and control system such that the missions are achieved.

A satellite system can be classified into several subsystems, including ADC (Attitude Determination and Control), OBDH (On Board Data Handling), TTC (Tracking, Telemetry and Command), Payload, etc. Each subsystem has its own function. For example, a function of the ADC subsystem is to control the satellite orientation such that a camera payload on the satellite can point toward the Earth or to a desired object.

Gyroscope or gyro sensor is a part of ADC subsystem. It is an inertial sensor that measures inertial angular motion of satellite expressed in a satellite body reference frame. Through a certain algorithm, combination of measurement value from gyro sensor and, for example, star sensor can determine the satellite attitude. Ivanov et al [5] presents Kalman filter based attitude determination with various combination of attitude sensors. This attitude information is not only required for attitude control need, but also to analyze mission results data. Therefore, for example, science data from SWUSV will be stored with the satellite attitude information [1].

Meanwhile, in a satellite development phases, component level verification is required via unit test [6]. Gyro sensor is one of the component that its functional operation must be verified. This paper proposed a testing tool to verify that a type of gyro sensor is functionally operating. To the best of author's knowledge, there is no existing article that presents the proposed testing tool.

The organization of this paper is as follows. In section II, background on the gyro type and study of some testing schemes that are covered as research methodology is presented. Drawback or problem in a gyro testing scheme is also defined. The proposed testing tool in order to avoid the drawback is presented in the section III. Experimental results and discussion regarding the proposed testing tool is written in section IV. This paper is covered with some concluding remarks including future research directions in section V.

## II. Methodology

### *A. Problem Definition and Motivation*

LITEF's gyro sensor has been used for many space application. Modular Optoelectronic Multispectral Scanner MOMS-2P—that was launched in April 1996 to be attached to the remote sensing module PRIRODA of the Russian space station MIR—has navigation system MOMS-NAV employing LITEF's gyro sensor [7][8]. In LAPAN-TUBSAT micro-satellite [9] as well as in DLR-TUBSAT and MAROC-TUBSAT micro-satellites [10], single-axis fiber optic µFORS-6 LITEF's gyro is used in pairs with an actuator called reaction wheel. The fiber optic gyro sensor regarded in this research is µFORS-4 LITEF's gyro, another series of µFORS LITEF's gyro.

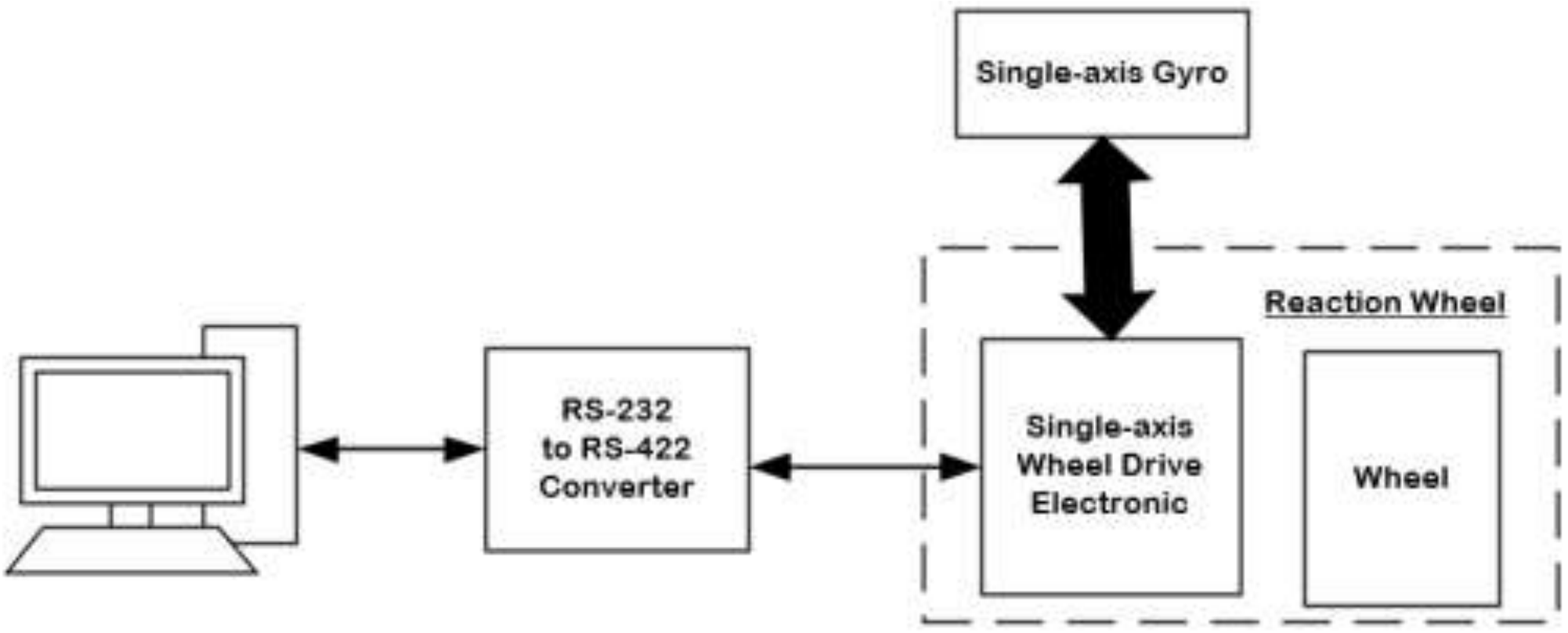

Figure 1. Function test bench scheme that utilizing wheel drive electronics of the reaction wheel

The µFORS-4 LITEF's gyro sensor comes with angle increment measurement mode, i.e. returns angle value of the gyro relative to the last data output. The returned data is a four bytes of data block: the first two bytes represents *measured angle data*, followed by one byte of *status* and the last one byte is *checksum* (Figure 2) — note that pre-calculation is needed to convert the first two byte into angle value in degree where beyond the scope of this paper. The data block is transmitted via serial transmission that compatible to RS-422 after requested by a trigger that follows *hardware trigger* type—one may refer to [11] for more comprehensive information about RS-422 standard. In contrast to the software trigger type that its request trigger signal is defined through two bytes data block (i.e. one byte of command followed by one byte of checksum), a request trigger signal of the hardware trigger type is formed by two discrete signals with particular timing diagram including interval of request frequency and delay of returned data after trigger.

| Byte No. 1 | Byte No. 2 | Byte No. 3 | Byte No. 4 |
|---|---|---|---|
| Most Significant Byte (Angle data) | Least Significant Byte (Angle data) | Status Byte | Checksum Byte |

Figure 2. Structure of a The Return Data

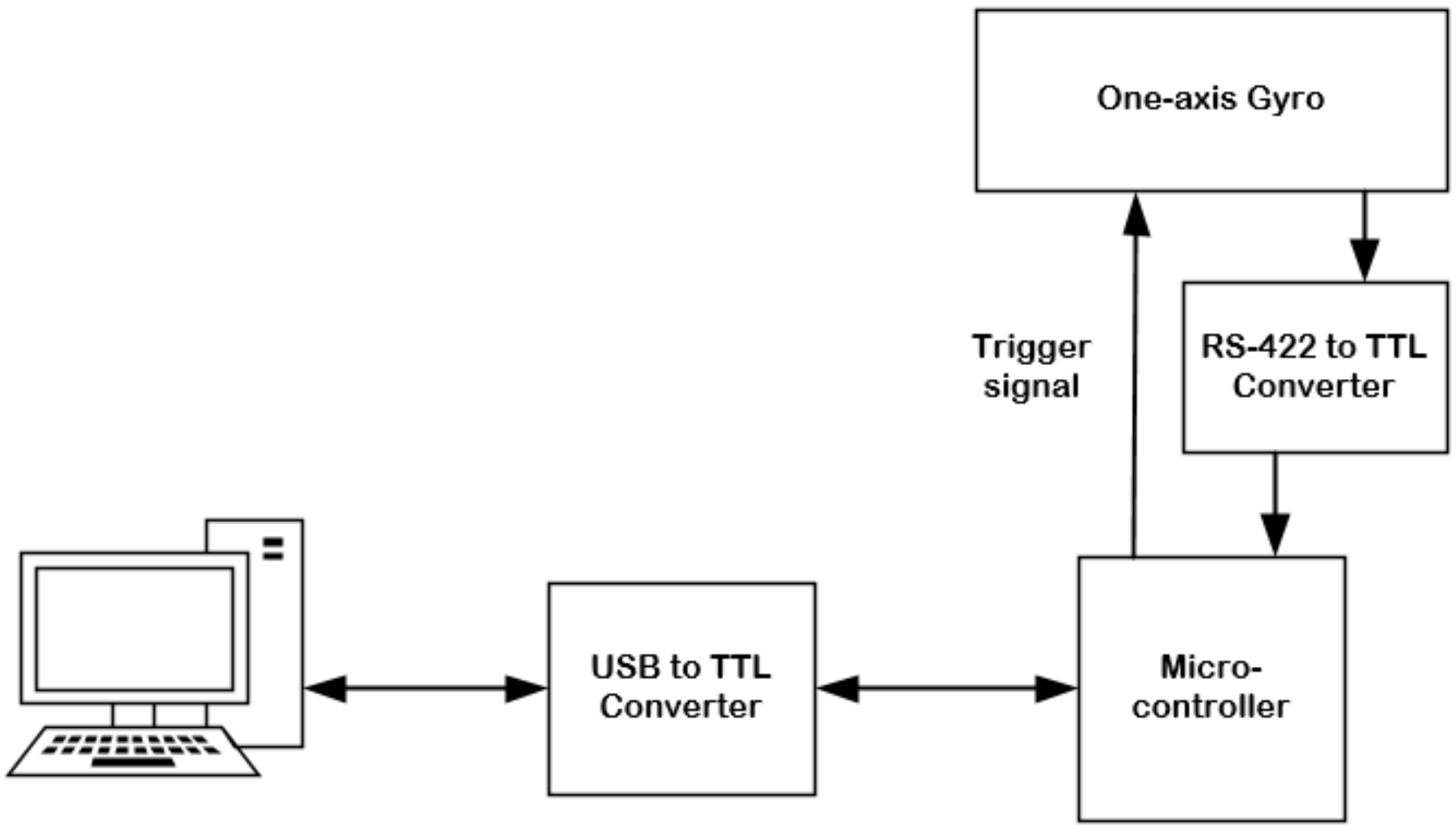

Figure 3. Alternative function test bench scheme that not utilizing wheel drive electronics

Since the gyro sensor is designed to work in pairs with reaction wheel [9] [10], functional testing of the gyro sensor is usually done by utilizing wheel drive electronics of the reaction wheel; it is illustrated in Figure 1. In such a testing scheme, there is software application run in the personal computer (PC) that simulates command signal from OBDH of satellite—one may refer to Figure 5 in reference [10] to acquire more information about the OBDH subsystem. Obviously, this testing scheme will be rely on another satellite component, i.e. the wheel drive electronics. Therefore, since the testing line is not directly connected from PC to the gyro, it comes with drawback that, if there is s failure between the gyro sensor and the wheel drive electronics, then source of problem may be hard to be identified.

Alternative test bench scheme is presented in Figure 3. In this scheme, microcontroller run firmware such that trigger signal can be generated and receives the returned data from gyro. This scheme reduces its dependence to the wheel drive electronics. However, the testing line is also not directly to the gyro, but via microcontroller. Therefore, how to build the test bench that independent to another satellite component as well as has direct connection line from PC to the gyro is challenging.

### B. *Hypothesis and Method*

In according to the gyro datasheet, request trigger signal of the hardware trigger type is in form of two discrete signals that follows the RS-422 standard. Then, its natural hypothesis is that it may possible to design a testing scheme that has direct connection line from PC to the gyro through serial communication satisfies RS-422 standard. It means that the request trigger signal of the hardware trigger type gyro is represented by particular data transmitted through serial communication from PC. Therefore, relationship between a data value transmitted from computer and its corresponding returned value from the gyro must be identified. Afterward, programming algorithm for generating the request trigger signal is formulated through some experiments.

## III. Result and Discussion

### A. *The Proposed Testing Tool Scheme*

Figure 4 presents the proposed testing tool scheme that independent to another satellite component and has direct connection line from PC to the gyro. In this scheme, the request trigger signal and the returned gyro data will be transmitted via USB to RS-422 converter cable—note that the voltage levels of the trigger signals and the returned data are CMOS levels referenced to +5 V that compatible to RS-422. In addition, to illustrate an alternative of testing and analysis flow, Figure 5 is presented. The trigger signal and returned gyro data will be handled by a desktop application. One may develop the desktop application in form of console application or graphical user interface (GUI) where data saving feature have to be accommodated for further analysis purposes. Therefore, the saved returned data can be inspected through an analysis tool, e.g. spread sheets, numerical analysis software or a self-developed application (programming software).

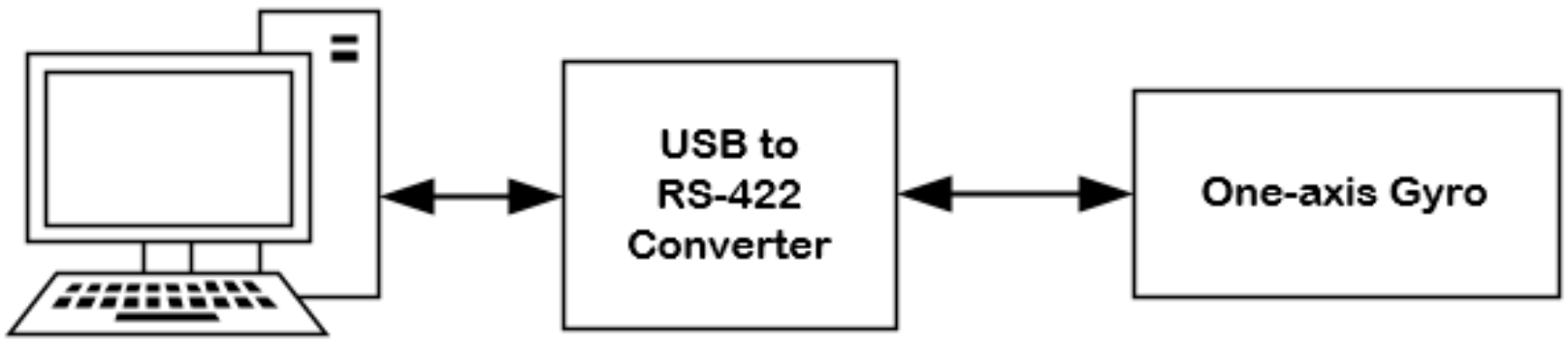

Figure 4. Scheme of the proposed testing tool

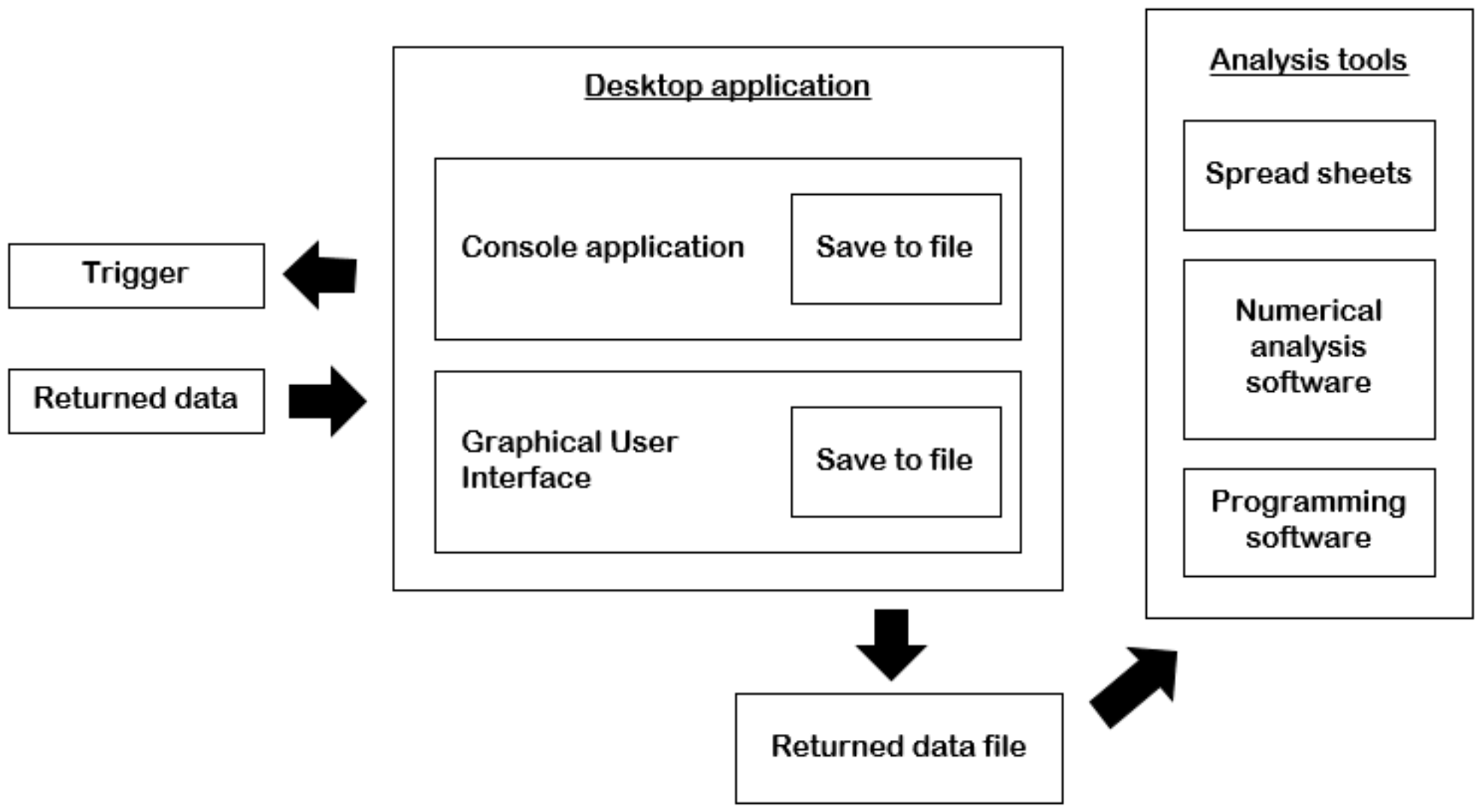

Figure 5. Testing and analysis flow

*B. Implementation and Analysis*

Figure 6 depicts implemented serial communication module between computer to the gyro and the gyro power source from computer. In the experiment, a desktop application in form of console application is developed. The application is developed in Visual Basic .NET programming language. Picture of the developed testing system for the experiment is shown in Figure 7. The experiment is done follows the following conditions:

- No intentional rotation is applied to the gyro sensor
- Algorithm of the developed application program is presented in **Error! Reference source not found.**;
- Experiment is run for all of one byte or 8-bit possible data values;
- Number of request trigger data transmission for each value is 10 times;
- Experiment is run for three different Sleep value setting, i.e. 80, 100 and 150 millisecond;
- Each saved returned data is opened in a spread sheets to be analyzed.

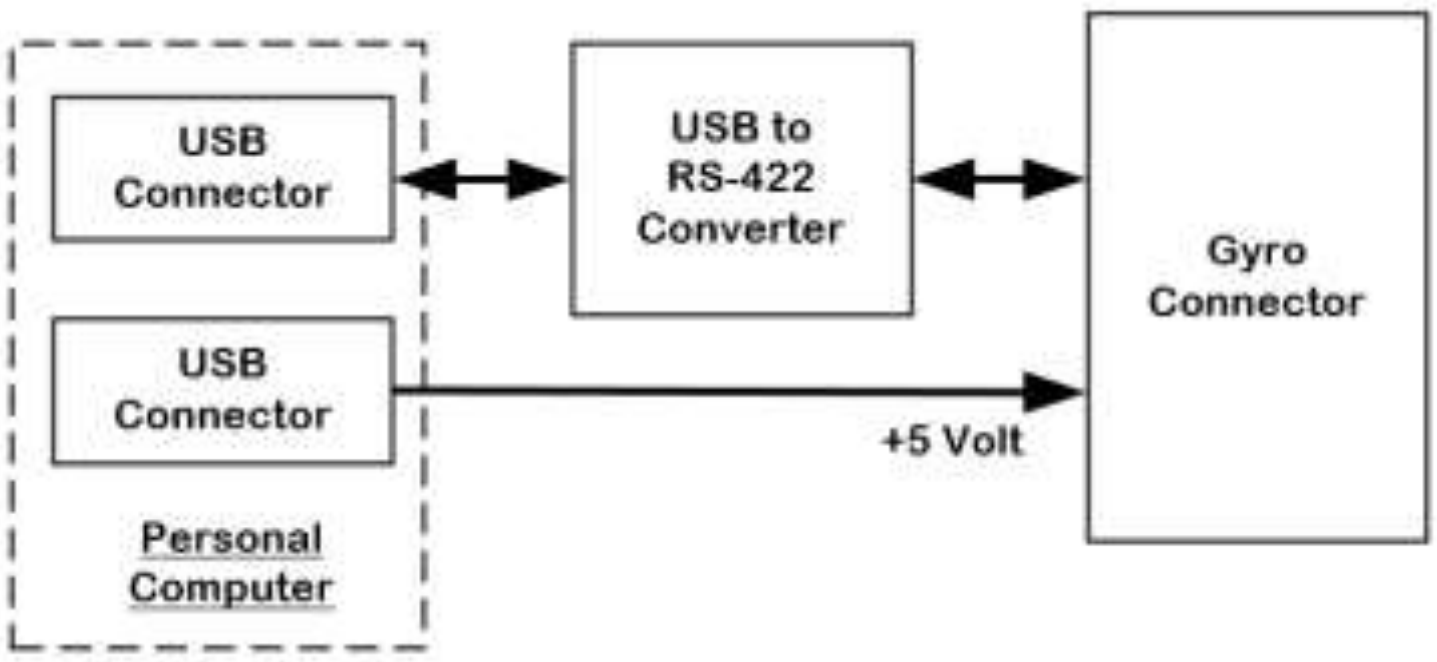

Figure 6. Scheme of the hardware part of the proposed testing tool

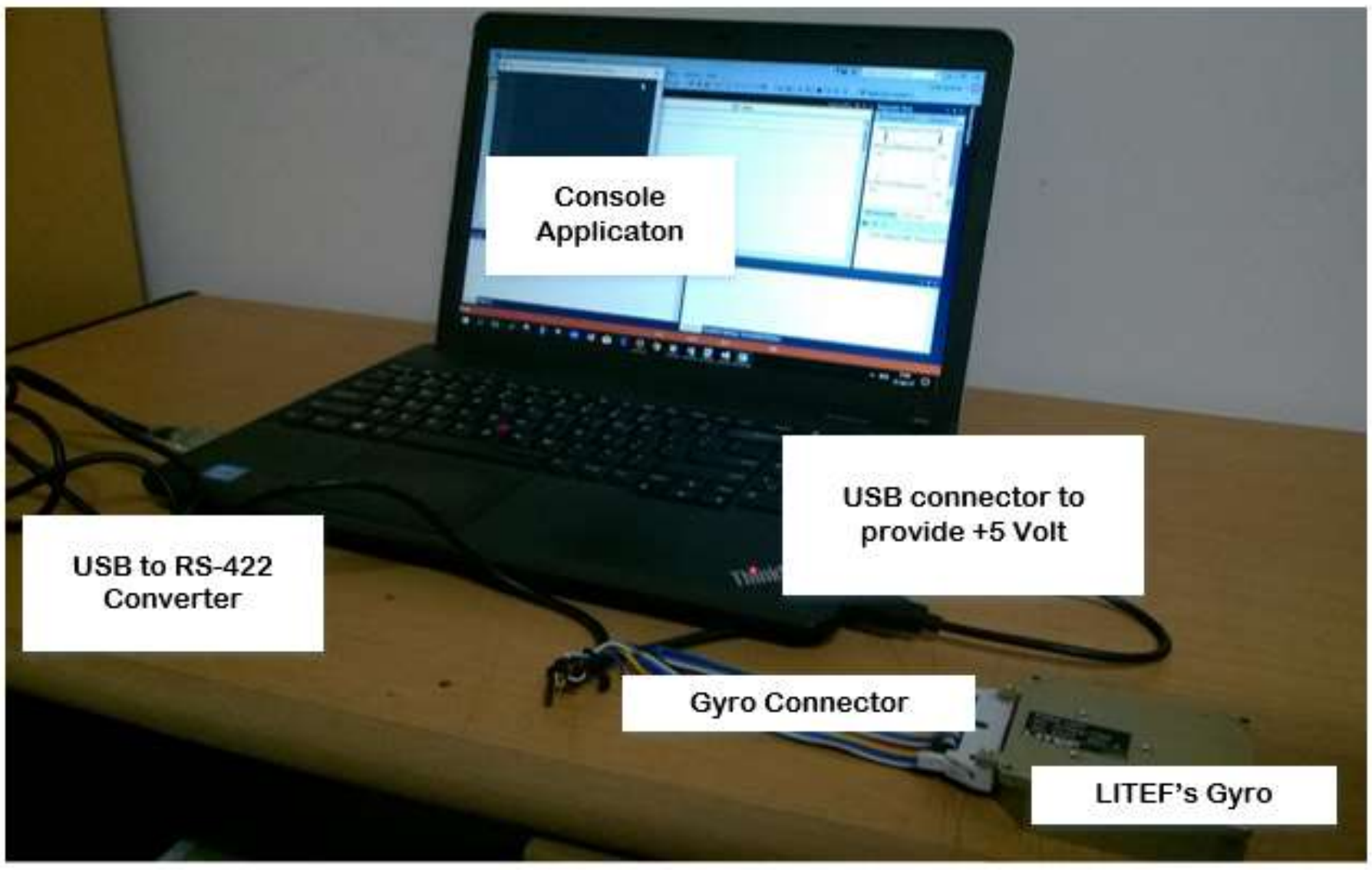

Figure 7. The developed testing tool

TABLE I. IMPLEMENTED ALGORITHM

| **Module Testing** |
|---|
| Declaration of variables |
| Main() |
| Initialization of serial communication |
| Initialization of byte data to be sent |
| Declare handler of serial port DataReceived event |
| z = 0 |
| While z < number of data transmitting |
| Write/ byte data to serial port |
| Sleep in milliseconds |
| Get bytes array of sumData |
| Write/ append the index 0 of byte array of sumData |
| sumData = 0 |
| End While |
| End Main |
| DataReceived event handler() |
| Gets the number of bytes to read |
| If bytes to read > 0 Then |
| For i = 0 To the number of bytes to read |
| sumData += the byte data that has been converted to Int16 |
| Write/ append the byte data to file |
| Next |
| End If |
| z += 1 |
| End DataReceived event handler |

Through the experiments run, it is interesting to discover that there are four types of returned data packet regarding the request trigger data signal. Note that the packet of returned data may consist of one or more returned data blocks, where a returned data block composed of two bytes of *angle data*, one byte of *status* and one byte of *checksum*, as described in Section II. As presented in Figure 8, the experiment resulted four sets of transmitted data that each of them raise different type of returned data packet. $Set_a$ , $Set_b$ and $Set_c$ consist of data that will raise one, two and three returned data block(s), respectively. While $Set_b \cap Set_c$ consists of data that will raise two or three returned data blocks alternately. Another fact appeared in the experiment is that every earliest returned data block presented a same status byte that refers to data request frequency that is too slow or the measurement range is exceeded.

TABLE II. FOR LOOP MODIFICATION

| **DataReceived event handler()** |
|---|
| Gets the number of bytes to read |
| If bytes to read > 0 Then |
| For i = 0 To 4 |
| sumData += the byte data that has been converted to Int16 |
| Write or append the byte data to file |
| Next |
| End If |
| z += 1 |

Afterward, since it is important to control a returned data by a trigger signal, hence the experiment for every data or element of $Set_a$ is re-run where the number of trigger transmission for each element is 1000 times. It is observed that, for Sleep value setting of 80 millisecond, at most 1 of 1000 of the returned data is in form of two returned data block. It is not experienced for 100 and 150 millisecond Sleep value setting. Fortunately, it is still possible to use 80 millisecond Sleep value setting by editing the loop For code inside the DataReceived event handler function, i.e. "the number of bytes to read" become "4" (**Error! Reference source not found.**). Fortunately, regardless property of control a returned data by a trigger signal, this modification is also implying to do trigger with any 8-bit data.

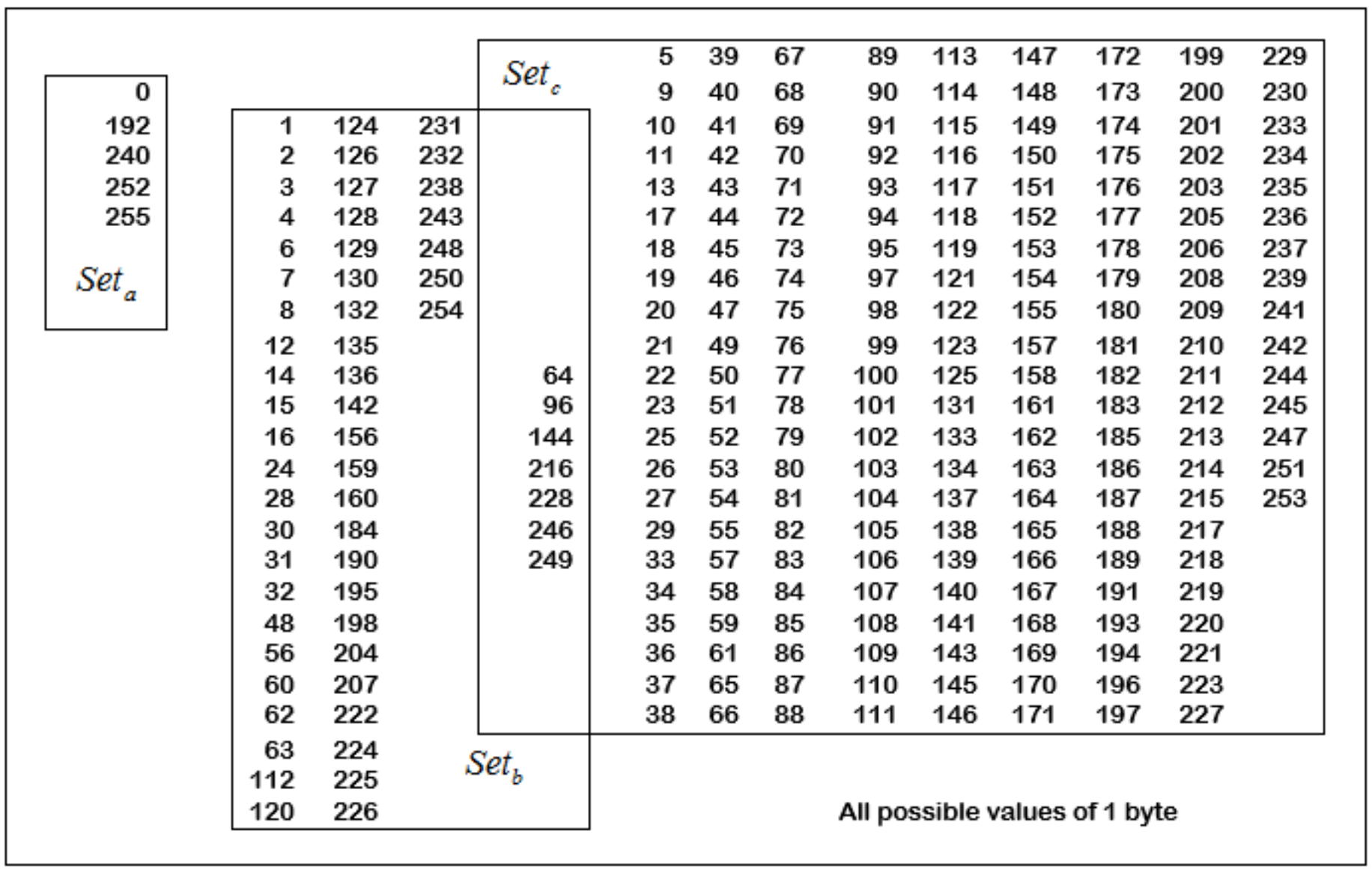

Figure 8. Set of data for the trigger signal

Beside the experiment described above, some Sleep values that smaller than 80 millisecond has been used, e.g. 1 millisecond and 40 millisecond. However, at least, there are two undesired results are experienced: the returned data is not received in form of a complete block or packet; and "IOException was unhandled" in relating to the process that cannot access the file where the returned data is to be saved.

At last, Figure 9 and Figure 10 are presented in order to completing the experiment results. Figure 9 presents the returned data file saving that implemented in the experiment. Meanwhile, Figure 10 presents plot of returned angle data obtained in the experiment after conversion from bytes value to degree value is applied.

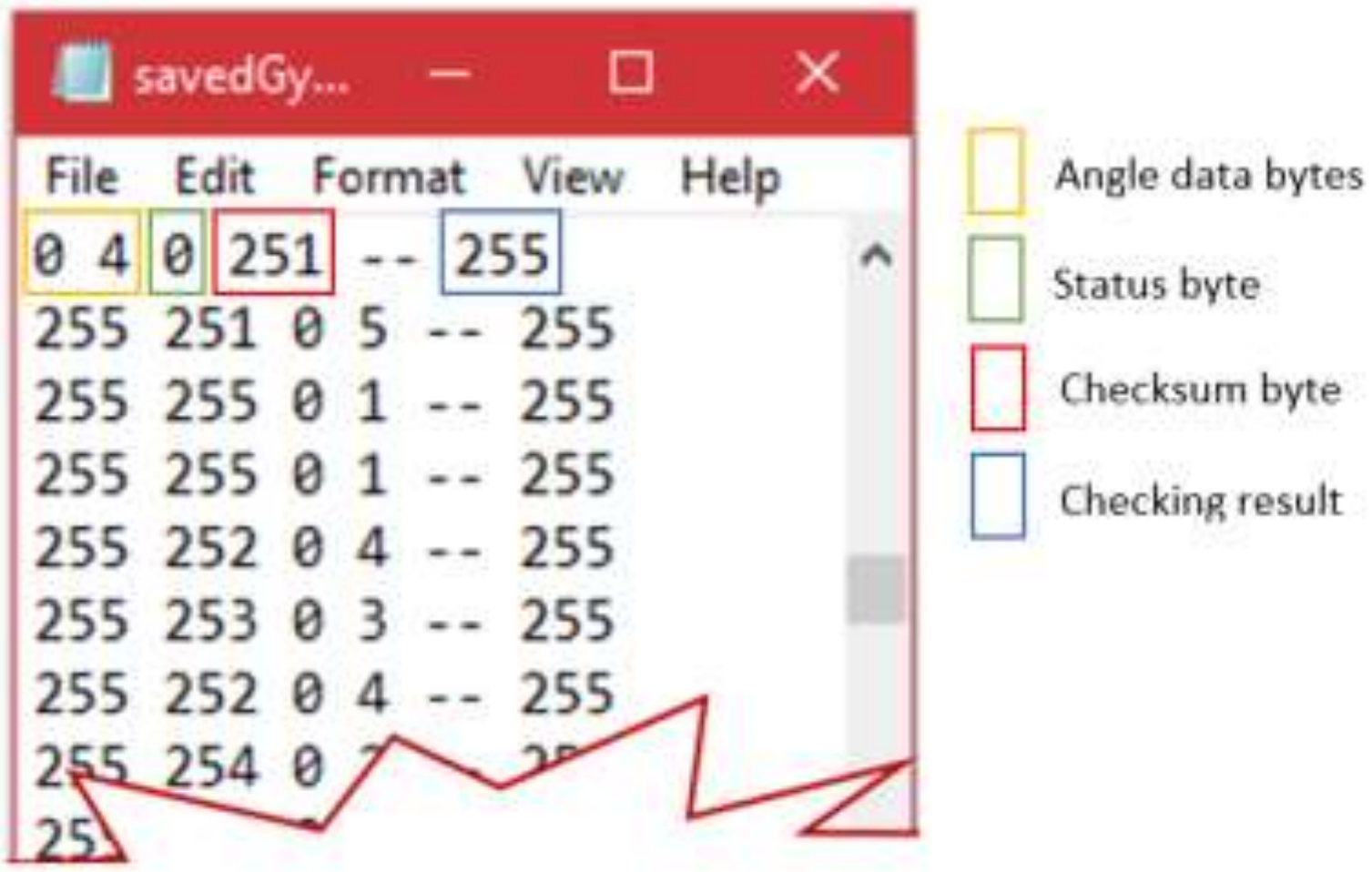

Figure 9. Returned data appearance in a saving file

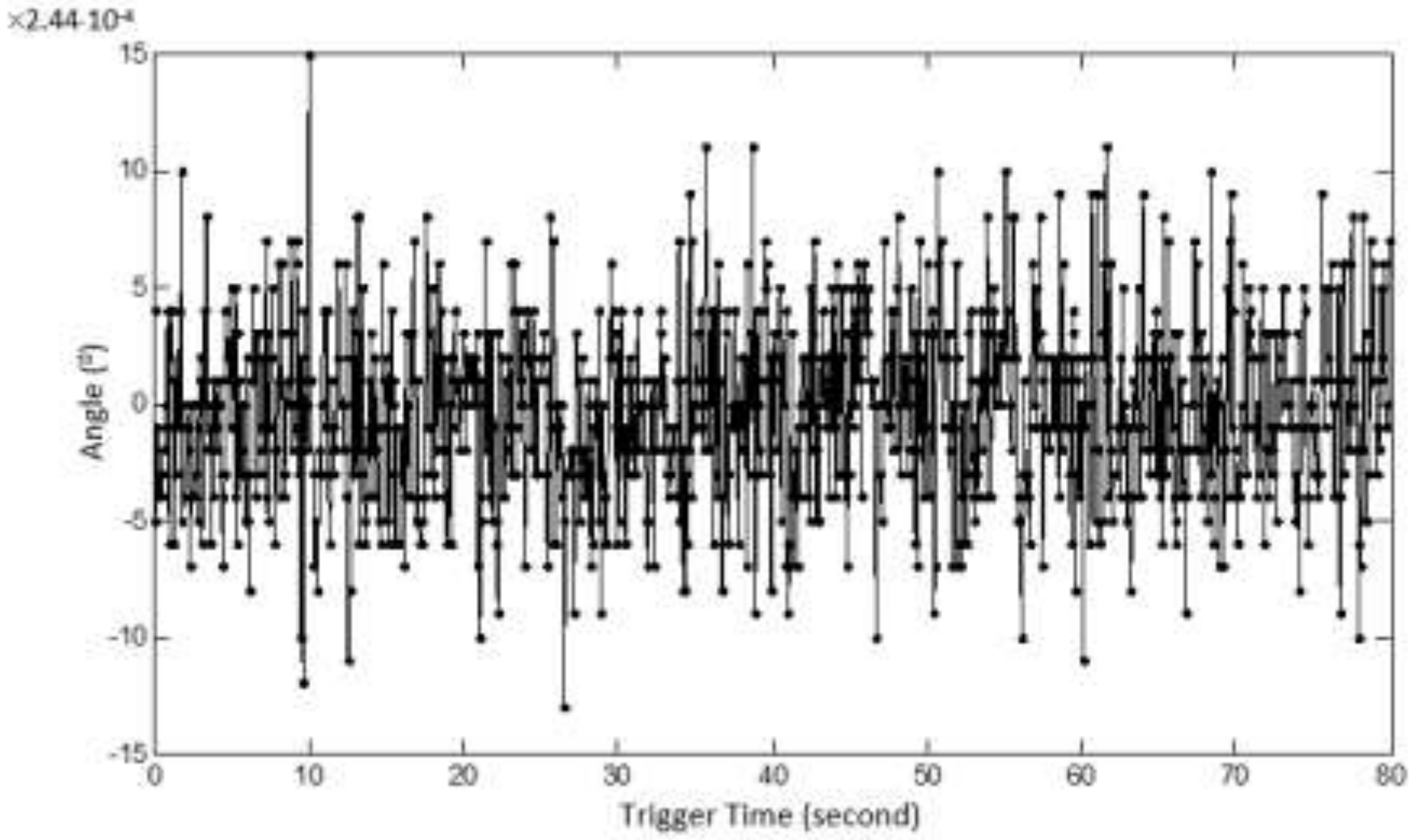

Figure 10. Plot of returned angle data obtained in the experiment

## IV. CONCLUDING REMARKS

The proposed testing tool for component level verification of a satellite gyro sensor consists of two main part, i.e. software (desktop application) and hardware (USB to RS-422 converter and USB cable for the gyro power supply). Three main feature must be accommodated in the desktop application are trigger transmitting, returned data receiving and returned data saving. USB to RS-422 converter is employed to transmit the trigger signal and to receive the returned gyro data (measurement data). Experiment according to the proposed testing tool has been conducted. Considering that to control returned data is essential, experimental results show that one trigger signal will raise one block of returned data if each 8-bit data of $Set_a$ is used in generating the trigger signal.

Future research may have several directions. First, one might be interested to focus on the design improvement that consist of man-computer interface aspect, e.g. graphical user interface, and integration of testing and analyzing into a single desktop application. Second, it is challenging to develop a hardware in the loop simulation based on this research result. For instance, to design an attitude determination system of satellite, it is promising to integrate the proposed scheme with star sensor simulator as reported in [12][13] as well as general satellite simulator as reported in [14], to name a few. Third, the proposed scheme may be applicable to be employed for earth mechanics related research, for example, Earth's rotary rate measurement [15].

## ACKNOWLEDGMENT

The authors would like to thanks to all members of 2016 Attitude Control System Experiment team. We also would like to acknowledge the Satellite Technology Center LAPAN management in supporting this research.